\documentclass[longauth]{aa}

\usepackage{graphicx} 
\usepackage[fleqn]{amsmath}
\usepackage{array, amssymb}
\usepackage[utf8]{inputenc}
\usepackage[T1]{fontenc}   
\usepackage{url}           
\usepackage{booktabs}      
\usepackage{lmodern}
\usepackage[dvipsnames]{xcolor}

\usepackage[colorlinks=True, citecolor=Blue, urlcolor=blue, breaklinks=false]{hyperref}
\usepackage{orcidlink}

\usepackage{fontawesome5}

\usepackage{listings}
\definecolor{codegreen}{rgb}{0,0.6,0}
\definecolor{codegray}{rgb}{0.5,0.5,0.5}
\definecolor{codepurple}{rgb}{0.58,0,0.82}
\definecolor{backcolour}{rgb}{0.95,0.95,0.92}
\lstdefinestyle{pythonstyle}{
    language=Python,
    backgroundcolor=\color{backcolour},   
    commentstyle=\color{codegreen},
    columns=fullflexible,
    keywordstyle=\color{magenta},
    numberstyle=\tiny\color{codegray},
    stringstyle=\color{codepurple},
    basicstyle=\ttfamily\footnotesize,
    breakatwhitespace=false,         
    breaklines=false,                 
    captionpos=b,                    
    keepspaces=false,                 
    numbersep=5pt,                  
    showspaces=false,                
    showstringspaces=false,
    showtabs=false,                  
    tabsize=2
}

\newcommand{\codehref}[2]{%
  \href{#1}{\textcolor{blue}{\texttt{#2}}}%
}

\makeatletter
\newcommand{\github}[1]{%
   \href{#1}{\faGithub}%
}
\makeatother

\usepackage{xspace}
\newcommand{\gray}{$\gamma$-ray\xspace}
\newcommand{\grays}{$\gamma$-rays\xspace}
\newcommand{\rxj}{RX\,J1713.7$-$3946\xspace}
\newcommand{\fermi}{\textit{Fermi}-LAT\xspace}
\newcommand{\hess}{H.E.S.S.\xspace}
\newcommand{\cmsTeV}{\,cm$^{-2}$\,s$^{-1}$\,TeV$^{-1}$\xspace}

\begin{document} 

  \title{Gammapy v2.0: \\ An upgrade of the Python package for gamma-ray astronomy}

  \author{
\texttt{Gammapy} team\thanks{Mail inquiries to:
        \href{mailto:GAMMAPY-COORDINATION-L@IN2P3.FR}{GAMMAPY-COORDINATION-L@IN2P3.FR}}:
   K.~Feijen\inst{\ref{APC}}\orcidlink{0000-0003-1476-3714}
   \and Q.~Remy\inst{\ref{IRAP}, \ref{MPIK}}\orcidlink{0000-0002-8815-6530}
   \and A.~Sinha\inst{\ref{tata},\ref{APC}}\orcidlink{0000-0002-9238-7163}
   \and R.~Terrier\inst{\ref{APC}}\orcidlink{0000-0002-8219-4667}
   \and A.~Donath\inst{\ref{CfA}}\orcidlink{0000-0003-4568-7005}
   \and F.~Acero\inst{\ref{CEA},\ref{IAC}}\orcidlink{0000-0002-6606-2816}
   \and T.~Bylund\inst{\ref{LUX}}\orcidlink{0000-0003-2946-1313}
   \and C.~Galelli\inst{\ref{INFN}}\orcidlink{0000-0002-7372-9703}
   \and B.~Khélifi\inst{\ref{APC}}\orcidlink{0000-0001-6876-5577}
   \and D.~Berge\inst{\ref{DESY},\ref{HUB}}\orcidlink{0000-0002-2918-1824} 
   \and C.~Boisson\inst{\ref{LUX}}\orcidlink{0000-0001-5893-1797}
   \and J.~L.~Contreras\inst{\ref{IPARCOS}}
   \and S.~Funk\inst{\ref{ECAP}}\orcidlink{0000-0002-2012-0080}
   \and M.~Fü{\ss}ling\inst{\ref{CTAO}}
   \and J.~Hinton\inst{\ref{MPIK}}
   \and R.~L\'opez-Coto\inst{\ref{IAA}}\orcidlink{0000-0002-3882-9477}
   \and F.~Pintore\inst{\ref{INAF}}\orcidlink{0000-0002-3869-2925}
   \and E.~de~O\~na~Wilhelmi\inst{\ref{DESY}}\orcidlink{0000-0002-5401-0744} 
   \and C.~{van Eldik}\inst{\ref{ECAP}}\orcidlink{0000-0001-9669-645X}
   \and R.~Zanin\inst{\ref{CTAO_RZ_int}}
   \and A.~Aguasca-Cabot\inst{\ref{UB}}\orcidlink{0000-0001-8816-4920}
   \and T.~Arnesen\inst{\ref{IAC-ULL}}\orcidlink{0009-0004-0816-0700}
   \and L.~Barrios-Jimenez\inst{\ref{IAC-ULL}}\orcidlink{0009-0008-6006-175X}
   \and J.~Bernete\inst{\ref{CIEMAT}}\orcidlink{0000-0002-8108-7552}
   \and N.~Biederbeck\inst{\ref{Dortmund}}\orcidlink{0000-0003-3708-9785}
   \and J.~Buchner\inst{\ref{MPE}}\orcidlink{0000-0003-0426-6634}
   \and J.~Djuvsland\inst{\ref{Bergen},\ref{MPIK}}\orcidlink{0000-0002-6488-8219}
   \and K.~Egg\inst{\ref{ECAP}}\orcidlink{0009-0002-4238-034X}
   \and G.~Emery\inst{\ref{IAA}}\orcidlink{0000-0001-6155-4742}
   \and S.~Fröse\inst{\ref{Dortmund}}\orcidlink{0000-0003-1832-4129}
   \and L.~Gréaux\inst{\ref{Bochum}}\orcidlink{0000-0003-0627-8436}
   \and J.~Konrad\inst{\ref{Dortmund}}\orcidlink{0009-0000-1257-4771}
   \and P.~Kornecki\inst{\ref{IAA}}\orcidlink{0000-0002-2706-7438}
   \and M.~Linhoff\inst{\ref{Dortmund}, \ref{CTAO}}\orcidlink{0000-0001-7993-8189}
   \and M.~Lundy\inst{\ref{McGill}}\orcidlink{0000-0003-3802-1619}
   \and K.~McKee\inst{\ref{Chicago}}\orcidlink{0000-0002-8547-8489}
   \and S.~Mender\inst{\ref{Dortmund}}\orcidlink{0000-0002-0755-0609}
   \and E.~Molina\inst{\ref{IAC-ULL}}\orcidlink{0000-0003-1204-5516}
   \and L.~Mohrmann\inst{\ref{MPIK}}\orcidlink{0000-0002-9667-8654}
   \and D.~Morcuende\inst{\ref{CTAO}}\orcidlink{0000-0001-9400-0922}
   \and M.~Nievas Rosillo\inst{\ref{IAC-ULL}}\orcidlink{0000-0002-8321-9168}
   \and L.~Olivera-Nieto\inst{\ref{API},\ref{GRAPPA}}\orcidlink{0000-0002-9105-0518}
   \and S.~Panny\inst{\ref{Innsbruck}}\orcidlink{0000-0001-5770-3805}
   \and R.~D.~Parsons\inst{\ref{HU-Berlin}}\orcidlink{0000-0003-3457-9308}
   \and N.~Pigoux\inst{\ref{LUPM}}\orcidlink{0009-0001-0100-9846}
   \and M.~Peresano\inst{\ref{MPP}}\orcidlink{0000-0002-7537-7334}
   \and T.~Preis\inst{\ref{Innsbruck}}\orcidlink{0009-0001-7110-6764}
   \and M.~Punch\inst{\ref{APC}}\orcidlink{0000-0002-4710-2165}
   \and M.~Regeard\inst{\ref{APC}}\orcidlink{0000-0002-3844-6003}
   \and G.~Roellinghoff\inst{\ref{ECAP}}\orcidlink{0000-0002-9824-9597}
   \and L.~Schlegel\inst{\ref{APC}}\orcidlink{0009-0002-3619-951X}
   \and B.~M.~Sipőcz\inst{\ref{IPAC}}\orcidlink{0000-0002-3713-6337}
   \and H.~Stapel\inst{\ref{APC}}
   \and K.~Streil\inst{\ref{ECAP}}\orcidlink{0009-0005-7886-1825}
   \and G.~S.~Subramanya\inst{\ref{GSI}}
   \and M.~Takahashi\inst{\ref{Nagoya}}\orcidlink{0000-0002-0574-6018}
   \and T.~Unbehaun\inst{\ref{ECAP}}\orcidlink{0000-0002-7378-4024}
   \and S.~Wong\inst{\ref{McGill}}\orcidlink{0000-0002-2730-2733}
   \and P.~Yu\inst{\ref{APC}}
}

\institute{
    Université Paris Cité, CNRS, Astroparticule et Cosmologie, F-75013 Paris, France \label{APC}
    \and Univ Toulouse, CNES, CNRS, IRAP, Toulouse, France \label{IRAP}
    \and Max-Planck-Institut für Kernphysik, P.O. Box 103980, D 69029 Heidelberg, Germany \label{MPIK} 
    \and Dept. of High Energy Physics, Tata Institute of Fundamental Research, 400005 Mumbai, India \label{tata}
    \and Center for Astrophysics | Harvard \& Smithsonian, USA \label{CfA}
    \and Universit\'e Paris-Saclay, Universit\'e Paris Cit\'e, CEA, CNRS, AIM, 91191 Gif-sur-Yvette, France \label{CEA}
    \and FSLAC IRL 2009, CNRS/IAC, La Laguna, Tenerife, Spain \label{IAC}
    \and LUX, Observatoire de Paris, Université PSL, Sorbonne Université, CNRS, 92190 Meudon, France \label{LUX}
    \and Sezione di Milano, Istituto Nazionale di Fisica Nucleare, via Celoria 16, 20133, Milano, Italy \label{INFN}
   \and Deutsches Elektronen-Synchrotron (DESY), D-15738 Zeuthen, Germany \label{DESY}
   \and Institute of physics, Humboldt-University of Berlin, D-12489 Berlin, Germany \label{HUB}
   \and IPARCOS Institute and EMFTEL Department, Universidad Complutense de Madrid, E-28040 Madrid, Spain \label{IPARCOS} 
   \and Friedrich-Alexander-Universität Erlangen-Nürnberg, ECAP, Nikolaus-Fiebiger-Str. 2, 91058 Erlangen, Germany \label{ECAP}
   \and CTAO ERIC, Science Data Management Centre (SDMC), Platanenallee 6, 15735 Zeuthen, Germany \label{CTAO}
   \and Instituto de Astrofísica de Andalucía-CSIC (IAA-CSIC), Glorieta de la Astronomía s/n, 18008, Granada, Spain \label{IAA}   
   \and INAF/IASF Palermo, via Ugo La Malfa 153, I-90146 Palermo, Italy \label{INAF}
   \and CTAO ERIC, Headquarters, Via Piero Gobetti 93/3, 40129 Bologna, Italy \label{CTAO_RZ_int}   
    \and Departament de Física Quàntica i Astrofísica, Institut de Ciències del Cosmos, Universitat de Barcelona, IEEC-UB, Barcelona, Spain \label{UB}
   \and Instituto de Astrofísica de Canarias and Dpto. de Astrofísica, Universidad de La Laguna, C/ Vía Láctea s/n, E-38205 La Laguna, Tenerife, Spain \label{IAC-ULL}
   \and Centro de Investigaciones Energéticas Medioambientales y Tecnológicas (CIEMAT), Madrid, Spain \label{CIEMAT}
    \and Astroparticle Physics Group, TU Dortmund University, Germany \label{Dortmund}   
   \and Max Planck Institute for Extraterrestrial Physics, Giessenbachstrasse 1, 85748 Garching, Germany \label{MPE}
   \and University of Bergen, Norway \label{Bergen} 
   \and Fakultät für Physik \& Astronomie, Ruhr-Universität Bochum, D-44780 Bochum, Germany \label{Bochum} 
   \and Physics Department, McGill University, Montreal, QC H3A 2T8, Canada \label{McGill}   
   \and University of Chicago, USA \label{Chicago}   
    \and Anton Pannekoek Institute for Astronomy, University of Amsterdam, Science Park 904, 1098 XH Amsterdam, The Netherlands \label{API}   
   \and Gravitation and Astroparticle Physics Amsterdam Institute, University of Amsterdam, Science Park 904, 1098 XH Amsterdam, The Netherlands \label{GRAPPA}   
   \and Universität Innsbruck, Institut für Astro- und Teilchenphysik, Technikerstraße 25, 6020 Innsbruck, Austria \label{Innsbruck}
   \and Institut für Physik, Humboldt-Universität zu Berlin, D 12489 Berlin, Germany \label{HU-Berlin}
   \and Laboratoire Univers et Particules de Montpellier, CNRS, Université de Montpellier, Montpellier, France \label{LUPM}
   \and Max-Planck-Institut für Physik, Boltzmannstraße 8, 85748 Garching bei München, Germany \label{MPP}
   \and Caltech/IPAC, USA \label{IPAC}
   \and GSI Helmholtzzentrum für Schwerionenforschung, Planckstraße 1 64291 Darmstadt, Germany \label{GSI}
   \and Institute for Space--Earth Environmental Research, Nagoya University, Furo-cho, Chikusa-ku, Nagoya 464-8601, Japan \label{Nagoya}
}

  \date{Received 14 July 2026}

  \abstract
   {Community-driven open-source software is becoming increasingly central to modern astronomy, enabling reproducible and transparent high-level science analysis across multiple instruments and collaborations. In this effort, \texttt{Gammapy}, the open-source \texttt{Python} package for \gray astronomy, plays a key role, as it provides a unified analysis framework accessible to the high-energy astrophysics community.}
   {In this paper, we present the second long-term release of \texttt{Gammapy}, v2.0, which introduces major statistical and methodological improvements to the analysis framework compared to paper for \texttt{Gammapy}, v1.0.} 
   {The new features of \texttt{Gammapy} include a joint test statistic and combined significance map estimation over multiple datasets, improved spectral uncertainty estimation via parameter sampling, Bayesian inference through nested sampling, a generalised sensitivity estimation framework, an energy-dependent morphology estimator, and new timing analysis utilities. Improvements have also been made in the documentation and the developer operations.}
   {As examples of the newly implemented features, we show combined significance maps from \fermi and \hess observations for \rxj and a HAWC sensitivity for an extended source with multiple event types. We perform a comparison of the `butterfly plots' between the previous version and current implementation and provide detailed discussions on upper limit calculations for very weak or non-existent sources. The power of nested sampling in probing multi-modal distributions is compared with traditional fitting results using a gradient descent from the \texttt{minuit} minimiser.}
   {\texttt{Gammapy} v2.0 consolidates a unified, instrument-agnostic framework for current and next-generation \gray observatories within a common  analysis environment. Compared to the previous stable release, v2.0 brings significantly greater statistical rigour, equipping the community with more robust and trustworthy tools for high-level science analysis.}
   

   \maketitle

\section{Introduction}
\label{sec:introduction}

The past decades have seen a significant shift in the astronomy community towards community-driven open-source software development, enabled in part by the maturation of the scientific \texttt{Python} ecosystem and, in particular, the emergence of \texttt{Astropy} \citep{astropy_2022}. Community standards such as the \texttt{FITS} format as a common language for astronomical data and distributed version control through \texttt{git} have  enabled transparent, collaborative, and openly reviewed development. Moreover, open observatories and collaborations for multi-messenger activities have been replacing proprietary instrument-specific software. \texttt{Gammapy} \citep{gammapy_v1}, an open-source \texttt{Python} package for the analysis of \gray data, has been a key player in this regard. First released in 2015, it is now widely adopted by the \gray community as a high-level analysis tool.

\texttt{Gammapy} was designed to enable end-to-end scientific workflows, processing high-level data products to provide physically meaningful scientific results. Data enter the \texttt{Gammapy} framework in the form of data level three (DL3), where \gray-like events and their corresponding instrument response functions (IRFs) are stored in a standardised format defined by the gamma-astro-data-format \citep[GADF;][]{gadf}. These are reduced to data level four (DL4), where the IRFs are applied to produce binned counts data assigned to physical sky positions. Finally, at data level five (DL5), physically meaningful source quantities such as fluxes, spectra, and light curves are derived for scientific interpretation. Further details on the package structure and data levels are outlined in the first \texttt{Gammapy} paper \citep{gammapy_v1}. Statistical inference in \texttt{Gammapy} is primarily based on a binned forward-folding technique in which a physically motivated model is convolved with the IRFs and compared directly to the measured data via maximum likelihood estimation using Poissonian statistics. 

A key design principle of \texttt{Gammapy} is instrument agnosticism. As long as data are provided in the GADF standard format, they can be analysed within the same framework regardless of the detector technology. This enables support for a broad and growing range of \gray instruments.  Alongside being the science analysis tool (SAT) for the next generation Cherenkov Telescope Array Observatory (CTAO), it is also officially used within the MAGIC and \hess collaborations. \texttt{Gammapy} v2.0 adds support for VERITAS \citep{veritas_2023}, ASTRI \citep{astri_2022}, and LST-1 \citep{LST_2023}, and it extends support beyond Imaging Atmospheric Cherenkov Telescopes to \fermi \citep{Fermi_2009} and HAWC \citep{hawc_2017}, reflecting the rapid progress in data format standardisation across the community. 

This paper describes the major new features and improvements introduced in \texttt{Gammapy} v2.0, the second long-term stable release of the package and the project status. Readers new to \texttt{Gammapy} are referred to \cite{gammapy_v1} for a detailed description of the package architecture and core analysis concepts. The focus here is on the incremental developments since v1.0. We introduce the basic statistical framework in Sect~\ref{section:statistical_background}. In Sect.~\ref{section:joint_ts} we describe the new support for joint test statistic estimation and combined significance maps across multiple datasets. Section~\ref{section:error_prop} presents an improved approach to spectral uncertainty estimation based on parameter sampling. Section~\ref{section:sampling} introduces Bayesian inference via nested sampling, and in Sect.~\ref{section:upper_lims} we discuss the limitations of classical upper limit estimation and present Bayesian alternatives. In Sect.~\ref{section:sensitivity} we describe a generalised sensitivity estimation framework, and Section~\ref{section:additional_improvements} presents a summary of further improvements, including energy-dependent morphology estimation, timing utilities, hypothesis testing for nested models, and infrastructure developments. Finally, in Sect.~\ref{section:community_software} we discuss the growth and community development of the Gammapy project, emphasising the importance of the collaborative effort. This section shows that the software is maintained and developed by a large group of contributors, many of whom are primarily supported through other means.

\section{Statistical background}
\label{section:statistical_background}
In this section we summarise the main components of the default statistical framework used in \texttt{Gammapy} to perform model fitting and uncertainty estimation.
Briefly stated, \texttt{Gammapy}'s statistical treatment relies on frequentist statistics using the binned maximum likelihood approach.
This constituted the complete framework in \texttt{Gammapy} v1.0, while v2.0 introduced the possibility of using Bayesian statistics (see Sect.~\ref{section:sampling} for further detail). 
In both cases, forward folding is adopted. This means that a physical model and the reduced instrument responses are combined to calculate predicted model counts, which are then compared to the measured counts through a likelihood function.

The complete statistical model has a number of physical parameters of interest ($\phi$; e.g. source flux, position) as well as nuisance parameters ($\theta$; e.g. additional sources, astrophysical backgrounds). 
To determine the best-fit values for the parameters of interest, $\hat{\phi}$, the likelihood, $ \mathcal{L}(\phi, \theta)$, is defined and maximised over all parameters:
\begin{equation}
\hat{\phi} = \underset{\phi,\, \theta}{\arg\max} \log \mathcal{L}(\phi, \theta) \, ,
\end{equation}
where $\mathcal{L}(\phi, \theta)$ is the (Poisson) likelihood of the model.

The explicit form of the likelihood implemented in \texttt{Gammapy} depends on the method to estimate the background: either from observed data or a parametrised model.
If the background is modelled by a template, it is treated as a parameter without statistical uncertainty, and the resulting fit statistic is the Cash-statistic \citep[\codehref{https://docs.gammapy.org/2.0.2/api/gammapy.stats.cash.html}{cash},][]{cstat}.
In contrast, when the background is estimated from data, it is included in the likelihood as a separate nuisance parameter, $\theta$; this is described by the W-statistic (\codehref{https://docs.gammapy.org/2.0.2/api/gammapy.stats.wstat.html}{wstat}).

When a likelihood has been chosen and using either analytical or numerical means, it is possible to maximise the likelihood with respect to the nuisance parameters and use the resulting maximising values $\hat{\theta}$ in the `profiled' likelihood,
\begin{equation} 
\hat{\mathcal{L}}(\phi) = \mathcal{L}(\phi, \hat{\theta}) = \underset{\theta}{ \max} \log \mathcal{L}(\phi, \theta),
\end{equation}
for each dataset.

\texttt{Gammapy}'s fitting functionality and parameter estimation routines, such as the flux estimator, are based on this approach, with separate likelihoods for each dataset. This natively provides a mechanism to obtain joint estimates from statistically independent observations because the log likelihoods can simply be summed together before maximising: 
\begin{equation}
\hat{\phi}_{\text{joint}} = \underset{\phi}{\arg \max} \sum_i\ \log \hat{\mathcal{L}}_i(\phi),
\end{equation}
with $\hat{\mathcal{L}}_i(\phi)$ as the profile likelihood of dataset $i$ and $\hat{\theta_i}$ as the profiled nuisance parameters of that dataset for a given $\phi$.
Note that this approach implicitly requires nuisance parameters be different from one dataset to the other.

Once calculated, this total (joint) likelihood profile can be used to estimate the parameter uncertainties or upper limits by transforming a confidence interval into a likelihood difference using the Wilks theorem, which states that the likelihood ratio, $\lambda(\phi) =  -2 \left(\hat{\mathcal{L}}(\phi) - \hat{\mathcal{L}}(\hat{\phi)} \right)$, is asymptotically distributed as a $\chi^2$  distribution.
Alternatively, symmetric uncertainties on fitted parameters can be approximated directly from the Hessian matrix (the matrix of second derivatives of the likelihood) constructed during numerical minimisation.

This frequentist approach works well when parameters are well constrained and the likelihood surface is approximately parabolic near its maximum. However, these assumptions can fail when parameters are poorly constrained by the data or when a parameter of interest lies at or near a physical boundary. In such cases, the asymptotic $\chi^2$ approximation underlying the likelihood ratio test can break down, leading to unreliable uncertainty estimates and upper limit derivation (see Sect.~\ref{section:upper_lims}).

A natural extension that addresses these limitations is to incorporate prior knowledge about the model parameters through a prior distribution, $\pi(\phi, \theta)$. Priors can take a variety of functional forms depending on the information available. For example, Gaussian priors can be used for parameters constrained by previous observations, while uniform priors can be used for parameters that lie within a specific range, stabilising the optimisation near boundaries where the Hessian estimate becomes unreliable. Combining the prior with the likelihood through Bayes' theorem gives the posterior distribution:
\begin{equation}
P(\phi, \theta \mid D) \propto \mathcal{L}(D \mid \phi, \theta)\, \pi(\phi, \theta)  \, .
\end{equation}

This naturally leads to the maximum a posteriori (MAP) estimate,
\begin{equation}
\hat{\phi}_\mathrm{MAP} = \underset{\phi,\,\theta}{\arg\max} \left[ \log \mathcal{L}(D \mid \phi, \theta) + \log \pi(\phi, \theta) \right],
\end{equation}
which can be understood as a penalised likelihood where the prior distribution acts as a penalty term. This preserves the frequentist point-estimate framework, allowing results to be interpreted in terms of confidence intervals and hypothesis testing, for example, while incorporating prior information about parameters and placing constraints on the solution.

However, MAP only provides a single point estimate and not the full uncertainty on the parameters. A fully Bayesian approach instead characterises the complete posterior distribution, from which credible intervals and marginalised uncertainties are obtained by integrating over nuisance parameters. This naturally propagates all sources of uncertainty without relying on asymptotic approximations. In this case, the Bayesian evidence,
\begin{equation}
\label{eq:Z}
Z = \int \mathcal{L}(D \mid \phi, \theta) \, \pi(\phi, \theta) \, \mathrm{d}\phi \, \mathrm{d}\theta,
\end{equation}
can be used for a quantitative model comparison. Here, $D$ denotes the data, $\mathcal{L}(D \mid \phi, \theta)$ is the likelihood of obtaining the data for a given parameter set, and $\pi(\phi, \theta)$ denotes the prior probability density over the parameters. 
In addition the computation of the Bayesian evidence can provide the full posterior distribution as a by-product (discussed in Sect.~\ref{section:sampling} in more detail).

\section{Joint test statistic and combined significance maps}
\label{section:joint_ts}
Combining information from multiple datasets is often essential when analysing data obtained from different instruments and under different observation conditions or distinct event selections, with each characterised by their own IRFs. In such cases, a joint likelihood provides a statistically coherent framework for integrating such heterogeneous data while preserving their individual statistical properties. 

\begin{figure*}
\centering\includegraphics[width=\textwidth]{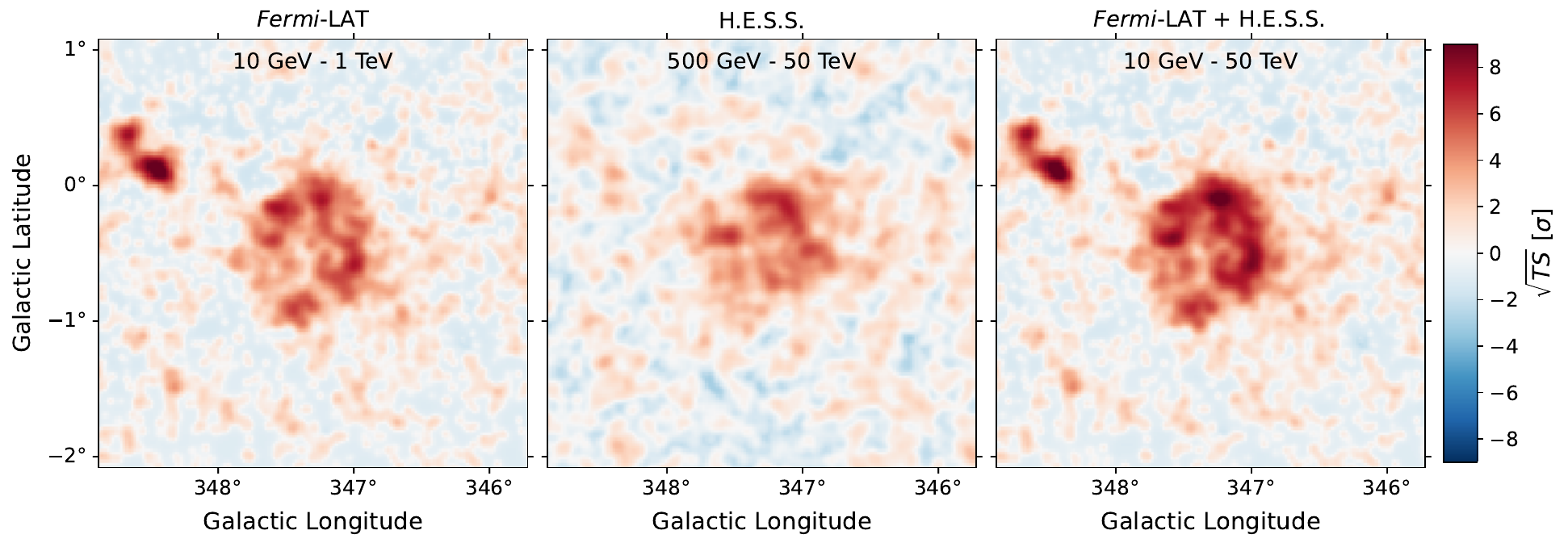}
\caption{Joint TS maps towards the \rxj region computed with the \codehref{https://docs.gammapy.org/2.0.2/api/gammapy.estimators.TSMapEstimator.html}{TSMapEstimator} class. Each panel corresponds to a different possible combination of datasets. The datasets used are \fermi PSF2 and PSF3 event types (left), 12 observations from H.E.S.S (middle), and \fermi plus H.E.S.S data (right). \github{https://github.com/gammapy/gammapy-2.0-paper/blob/main/scripts/joint/joint_maps_fermi_lat_hess.py}}
\label{fig:joint_ts} 
\end{figure*}

Joint likelihood maximisation for model parameter optimisation has been available since \texttt{Gammapy} v1.0. However, the estimation of joint test statistic (TS) maps was not previously implemented. The \codehref{https://docs.gammapy.org/2.0.2/api/gammapy.estimators.TSMapEstimator.html}{TSMapEstimator} computes a TS map by evaluating a test source at successive positions on the sky and comparing the likelihood of a model including this source to that of a background model at each position. No positivity constraint is applied to the source-model flux, and the sign of the TS follows that of the estimated flux. The result is stored as a \codehref{https://docs.gammapy.org/2.0.2/api/gammapy.estimators.FluxMaps.html}{FluxMaps} object, which provides access to all computed quantities.
With \texttt{Gammapy} v2.0, support for joint estimation over multiple datasets was introduced, i.e. the \texttt{run} method of an estimator can accept a \codehref{https://docs.gammapy.org/2.0.2/api/gammapy.datasets.Datasets.html}{Datasets} object, which is a list of binned datasets.
In this joint approach, the predicted counts in each dataset are computed while accounting for the different IRFs and background models. The likelihood contributions are summed to construct the joint likelihood at each sky position. This feature should only be used with statistically independent datasets, for example, different IRF types, different observations, or different instruments that do not record the same events.
We demonstrate the joint TS map computation in three representative scenarios towards the \rxj region: the combination of different \fermi event types (PSF2+PSF3), the combination of multiple \hess observations acquired at different times, and the combination of both the \fermi and \hess datasets. These examples are illustrated in Fig.~\ref{fig:joint_ts}, where the \hess dataset is taken from \cite{hgps_2018} also available in \codehref{https://github.com/gammapy/gammapy-data}{GAMMAPY\_DATA}.

\begin{figure*}
\centering\includegraphics[width=\textwidth]{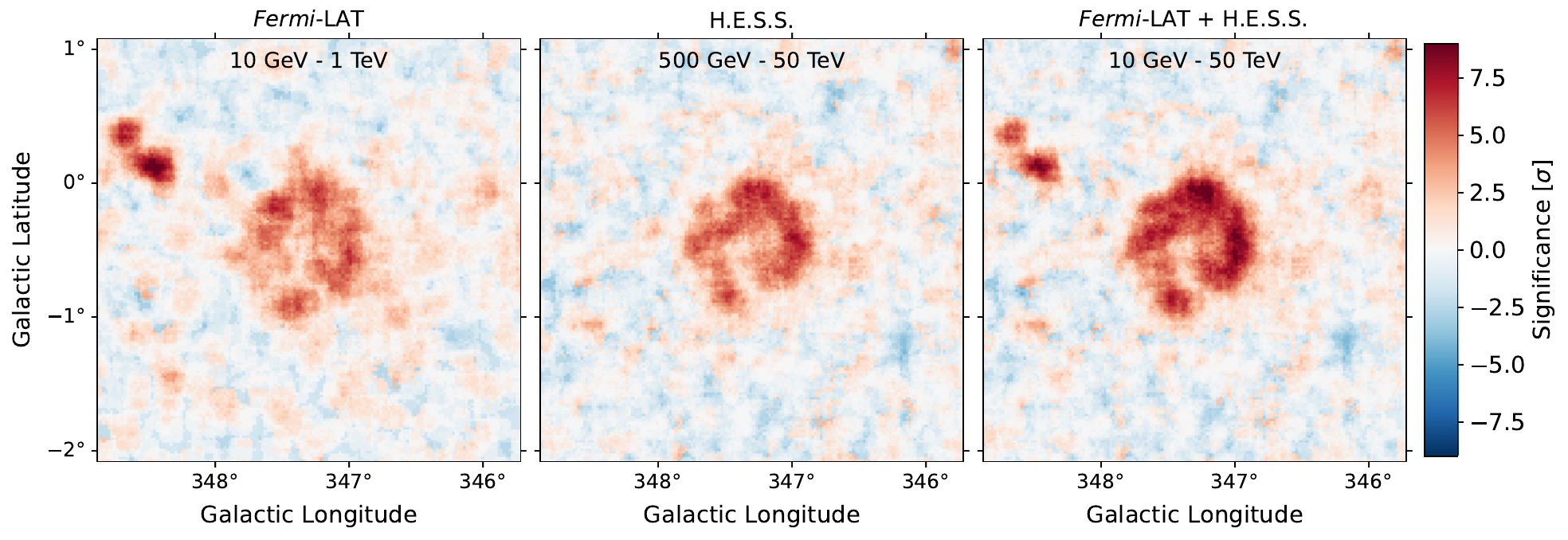}
\caption{Combined significance maps towards the \rxj region computed for a correlation radius of 0.1$^\circ$ with the \texttt{ExcessMapEstimator} class and \texttt{get\_combined\_significance\_maps} function. Each panel corresponds to a different possible combination of datasets. The datasets used are \fermi PSF2 and PSF3 event types (left), 12 observations from H.E.S.S (middle), and \fermi plus H.E.S.S (right). \github{https://github.com/gammapy/gammapy-2.0-paper/blob/main/scripts/joint/joint_maps_fermi_lat_hess.py}}
\label{fig:combined_significance}
\end{figure*}

The \codehref{https://docs.gammapy.org/2.0.2/api/gammapy.estimators.utils.combine_flux_maps.html}{combine\_flux\_maps} function can be used to combine \codehref{https://docs.gammapy.org/2.0.2/api/gammapy.estimators.FluxMaps.html}{FluxMaps} obtained from previous estimates on independent datasets. The combination can be performed by summing together likelihood profiles when available, i.e. the flux maps must contain a \texttt{stat\_scan} attribute, which stores the likelihood profile information as a spatial map for the scanned parameter. Otherwise, a likelihood profile approximation is computed with the available quantities, assuming that probability distributions for flux points correspond to Gaussians and those for upper limits to complementary error functions \citep[see Eq. C7 of ][]{2013ApJ...773..168M}. If asymmetric errors are available, asymmetric Gaussians distribution can be used.
Due to the data volume of the likelihood profiles, this functionality requires more memory than the direct joint estimate with the \codehref{https://docs.gammapy.org/2.0.2/api/gammapy.estimators.TSMapEstimator.html}{TSMapEstimator} class.

The main limitation of the joint and combined TS maps is that the spectral model of the test source is assumed to be valid across all datasets. If this hypothesis is relaxed, i.e imposing no constraint on the spectral shape, an alternative approach can be used to perform the combination a posteriori. The \codehref{https://docs.gammapy.org/2.0.2/api/gammapy.estimators.utils.combine_significance_maps.html}{combine\_significance\_maps} function computes a combined significance map following the formalism described by \cite{Lancaster1961} from a list of TS maps estimated over independent datasets. The TS maps are natively contained in the \codehref{https://docs.gammapy.org/2.0.2/api/gammapy.estimators.FluxMaps.html}{FluxMaps} objects created by the estimators. The combined significance estimation is based on the assumption that, for each valid energy bin in each dataset, the model introduces one degree of freedom and that the corresponding TS follows a $\chi^2$ distribution. Under these conditions, the TS values from independent energy bins and datasets can be summed, and the resulting total TS is itself $\chi^2$-distributed, with the number of degrees of freedom given by the sum over all contributing bins. The combined significance is then derived from this distribution, ensuring that the final result is correctly adjusted for the number of degrees of freedom. 

Figure~\ref{fig:combined_significance} illustrates three examples of combined significance maps towards the \rxj region computed with \fermi and \hess data.
The \codehref{https://docs.gammapy.org/2.0.2/api/gammapy.estimators.utils.get_combined_significance_maps.html}{get\_combined\_significance\_maps} function can also be used to run the estimator and to perform the combination over a list of datasets sequentially. An example code snippet illustrating the production of a combined significance map towards the Crab Nebula using HAWC Pass4 event types is shown in Fig.~\ref{code:combined_significance}.
This approach is faster and less memory-consuming, especially if the \codehref{https://docs.gammapy.org/2.0.2/api/gammapy.estimators.ExcessMapEstimator.html}{ExcessMapEstimator} is used instead of the \codehref{https://docs.gammapy.org/2.0.2/api/gammapy.estimators.TSMapEstimator.html}{TSMapEstimator}, as shown in the example.\footnote{It is important to note that the \codehref{https://docs.gammapy.org/2.0.2/api/gammapy.estimators.ExcessMapEstimator.html}{ExcessMapEstimator} internally uses a simple backward folding based on analytical solutions by \cite{LiMa1983}, whilst the \codehref{https://docs.gammapy.org/2.0.2/api/gammapy.estimators.TSMapEstimator.html}{TSMapEstimator} performs a full forward folding.}

\begin{figure}[h]
\begin{lstlisting}
from gammapy.datasets import Datasets
from gammapy.estimators import ExcessMapEstimator
from gammapy.estimators.utils import (
    get_combined_significance_maps
)

# read datasets containing different event types
path =  "$GAMMAPY_DATA/hawc/DL4/"
filename = f"{path}HAWC_pass4_public_Crab.yaml"
datasets = Datasets.read(filename)

# compute  combined signifiance map
estimator = ExcessMapEstimator(
    correlation_radius="0.2 deg"
)
combined = get_combined_significance_maps(
    estimator,
    datasets,
)
\end{lstlisting}
\caption{Computation of combined significance map towards the Crab Nebula over the different event types of the HAWC-pass4 public dataset.}
\label{code:combined_significance}
\end{figure}

The limitation of both of the methods outlined above is that the spatial geometry of the datasets must be aligned. In contrast, the energy axes do not necessarily have to be aligned or even overlap. 
A general limitation of combining different datasets to obtain significance or flux maps is the systematic differences present in the instrument responses. It is therefore important to account for these systematic effects that could make the combination unreliable. When performing joint modelling, some systematic effects should be included. One way to account for them is at the background-model level (\codehref{https://docs.gammapy.org/2.0.2/api/gammapy.modeling.models.FoVBackgroundModel.html}{FoVBackgroundModel}), where nuisance parameters can be introduced. Another broader example of how to include systematic effects is outlined in \cite{nigro_2019}.

\section{Error propagation and butterfly plots}
\label{section:error_prop}

Spectral band (also known as butterfly) plots are common in \gray astronomy. They indicate the propagated uncertainties of a fitted spectral model, for instance, in a spectral energy distribution (SED). 
In \texttt{Gammapy} v1.0, spectral bands were computed using classical error propagation theory. This approach relies on a first-order linearisation of the spectral model around the best-fit parameter values using the covariance matrix extracted at that position. This method is therefore only valid when the relative parameter uncertainties are small. This limitation becomes particularly severe when extrapolating the spectral model far from the de-correlation energy, where the linear approximation can produce misleading or even unphysical results.

\begin{figure}
\centering\includegraphics[width=\columnwidth]{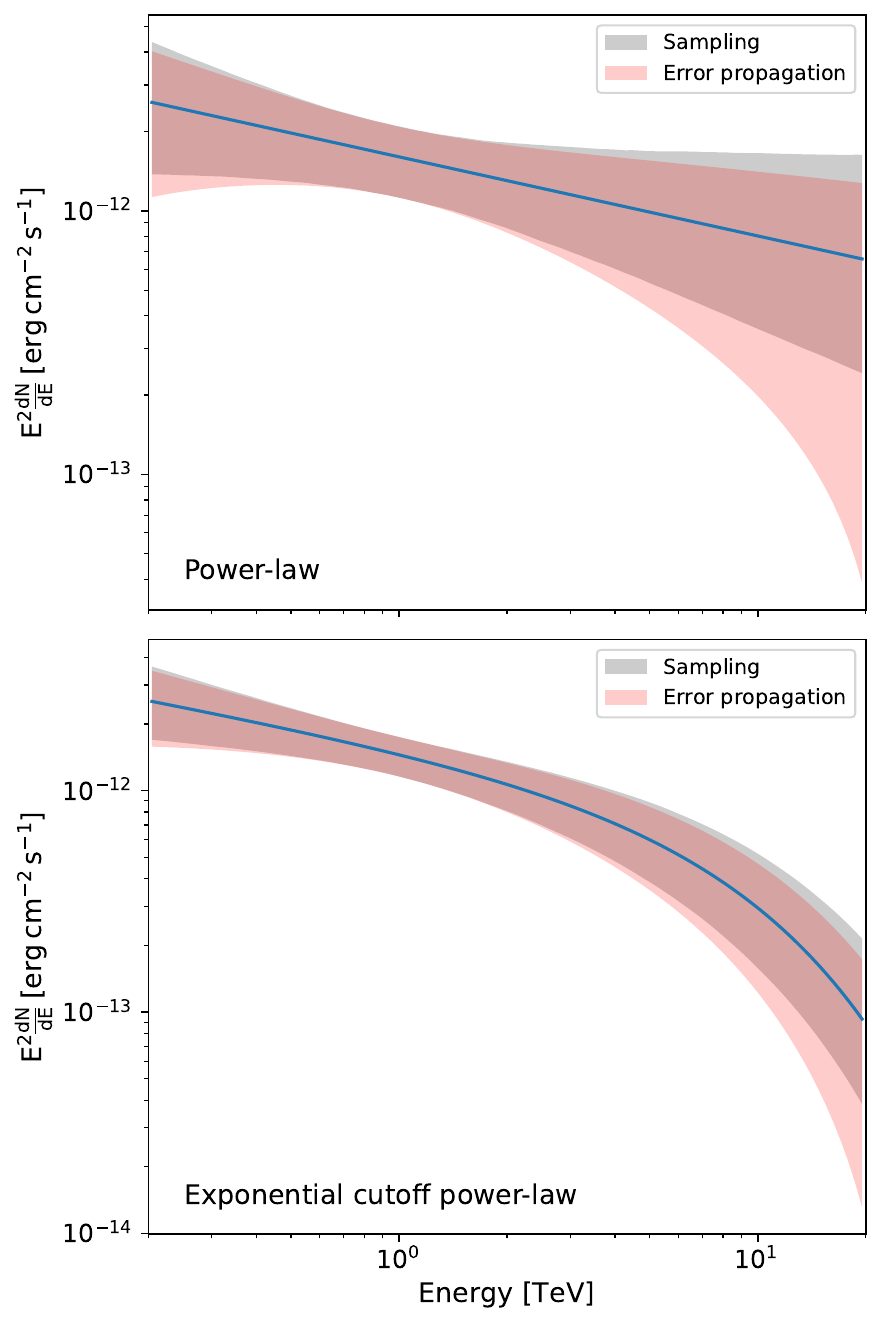}
\caption{Comparison of SED error bands (butterfly plots) produced with error propagation and parameter sampling for a power-law model (top) and an exponential cutoff power law (bottom). \github{https://github.com/gammapy/gammapy-2.0-paper/blob/main/scripts/butterfly.py}}
\label{fig:sampling_SED}
\end{figure}

To overcome this issue, \texttt{Gammapy} v2.0 computes an envelope of predicted SEDs for the fitted model using a sampling of its parameters. The parameter samples are generated from a multi-normal random variable using the best-fit values as the mean and the fit covariance matrix. At each energy, the quantiles containing 16\% and 84\% of the predicted energy flux are used as the envelope boundary values. A comparison of the predicted SEDs for a poorly constrained power law and a power law with an exponential cutoff spectral model is shown in Fig.~\ref{fig:sampling_SED}.

Thanks to the vectorisation of spectral model evaluations, this method still produces SEDs in an efficient manner. A few models cannot be evaluated using vectorisation,\footnote{Namely the \codehref{https://docs.gammapy.org/2.0.2/api/gammapy.modeling.models.NaimaSpectralModel.html}{NaimaSpectralModel}.} in which case multi-processing is used. The parameter sampling method now replaces error propagation in all spectral model error evaluation (integral and energy flux uncertainty, local spectral index uncertainty, etc.).

\section{Bayesian inference with nested Sampling}
\label{section:sampling}

Bayesian methods are increasingly used in high-energy astrophysics to infer model parameters and to compare competing physical scenarios. When performing model comparison, one can compute the Bayesian evidence $Z$ (as defined in Eq.~\ref{eq:Z}), which is the marginal likelihood of the model, and it quantifies how well the model as a whole predicts the data.

Traditional Markov chain Monte Carlo (MCMC) methods approximate the posterior distribution by evolving correlated sequences of samples (called `chains') through parameter space. While widely used, problems with their convergence to the true distribution can be difficult to diagnose. Furthermore, on complex likelihood surfaces, MCMC chains may miss some modes of the posterior distribution. Standard MCMC methods do not provide an estimate of the evidence $Z$ directly, requiring additional approximations for model comparison.

Nested sampling \citep{Skilling2004} addresses these limitations by transforming the multi-dimensional evidence integral (Eq.~\ref{eq:Z}) into a one-dimensional integral over the enclosed prior mass \citep[see][for recent reviews]{Ashton2022,Buchner2023}. The algorithm maintains a set of `live points' drawn from the prior. At each iteration, the point with the lowest likelihood is replaced by a new point constrained to have a higher likelihood than the discarded one. As iterations proceed, the allowed region of prior volume shrinks monotonically, naturally mapping the structure of the posterior while simultaneously building up an evidence estimate up to a user-defined remaining volume fraction. All live points explore the global prior space, making nested sampling less sensitive to parameter initialisation position and effective at identifying multi-modal and highly correlated posterior structures. Posterior samples are obtained as a by-product from the sequence of discarded points.

In \texttt{Gammapy} v2.0 we introduced the nested sampling implementation through the \codehref{https://docs.gammapy.org/2.0.2/api/gammapy.modeling.Sampler.html}{Sampler} interface with the \codehref{https://johannesbuchner.github.io/UltraNest/}{UltraNest} backend \citep{Buchner2021}. Once the priors for all free model parameters to be sampled have been defined and the sampler configuration set up, the \texttt{run} method can be called with a given \codehref{https://docs.gammapy.org/2.0.2/api/gammapy.datasets.Datasets.htm}{Datasets} object as an argument. A minimal example is shown in Fig. \ref{code:sampler}, where the technical settings typically required for the \codehref{https://docs.gammapy.org/2.0.2/api/gammapy.modeling.Sampler.html}{Sampler} are outlined.

\begin{figure}[h]
\begin{lstlisting}
from gammapy.modeling import Sampler
from gammapy.modeling.models import (
    PowerLawSpectralModel,
    UniformPrior,
    LogUniformPrior,
)
pl=PowerLawSpectralModel()
pl.amplitude.prior = LogUniformPrior(
    min=1e-12, 
    max=1e-10
)
pl.index.prior = UniformPrior(min=1, max=5)
model = SkyModel.create(spectral_model=pl)
datasets.models = [model]
sampler_opts = {
    "live_points": 400,
    "frac_remain": 0.1,
    "log_dir": None,
}
sampler = Sampler(backend="ultranest",
                  sampler_opts=sampler_opts)
result = sampler.run(datasets)
\end{lstlisting}
\caption{Running the \codehref{https://docs.gammapy.org/2.0.2/api/gammapy.modeling.Sampler.html}{Sampler} class with the \codehref{https://johannesbuchner.github.io/UltraNest/}{UltraNest} backend for given datasets. The model parameters to be sampled require defined priors. The \texttt{live\_points} parameter controls exploration; higher values better resolve multi-modal or narrow posterior features at increased computational cost. The \texttt{frac\_remain} parameter sets the stopping criterion via the fraction of posterior mass remaining in the live points. Lower values yield more accurate evidence estimates but longer runtimes. The \texttt{log\_dir} option enables diagnostic output, which is useful for long or interrupted computation runs.}
\label{code:sampler}
\end{figure}

An important aspect of Bayesian inference is the choice of priors. Nested sampling requires explicit priors for all free parameters, and the performance and interpretability of the results can be influenced by these choices. Parameters spanning several orders of magnitude (e.g. flux normalisations) should be assigned log-uniform priors (\codehref{https://docs.gammapy.org/2.0.2/api/gammapy.modeling.models.LogUniformPrior.html}{LogUniformPrior}), while parameters constrained to a finite range (e.g.\ spectral indices, spatial extents) are naturally handled with uniform priors (\codehref{https://docs.gammapy.org/2.0.2/api/gammapy.modeling.models.UniformPrior.html}{UniformPrior}). As the evidence explicitly depends on the priors, adopting physically motivated and well-justified prior ranges is essential when comparing alternative models.

Overall, nested sampling provides a robust framework for model inference, yielding both reliable posterior distributions and direct evidence values suitable for quantitative model comparison. Its implementation in \texttt{Gammapy} makes it straightforward to apply to \gray data while retaining flexibility through tunable sampler options, as illustrated in the code snippet in Fig.~\ref{code:sampler}.

\begin{figure}[hbt!]
\centering
\includegraphics[width=\columnwidth]{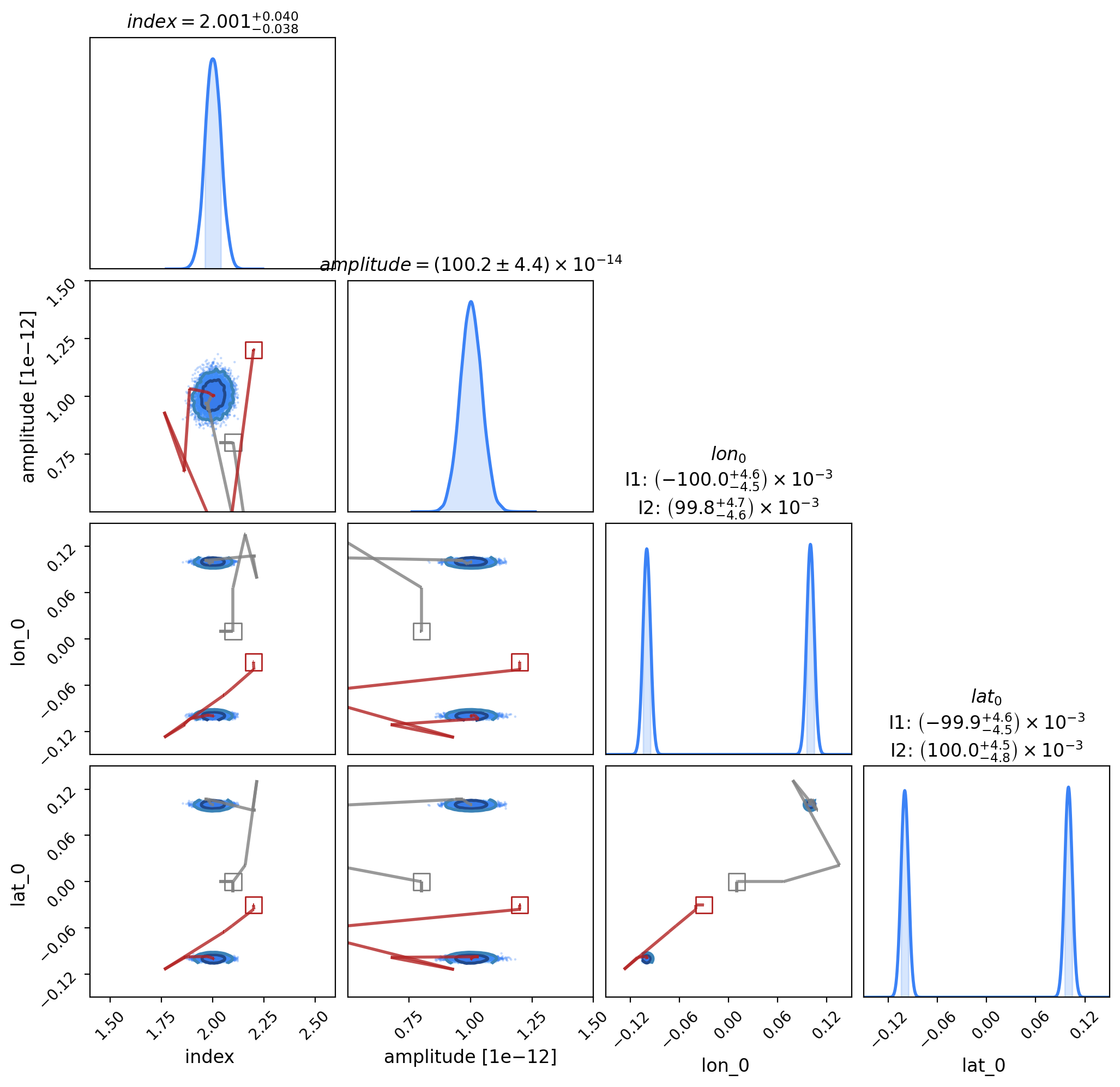}
\caption{Posterior distribution of the multi-modal  problem analysed with the nested sampling approach as compared to the gradient descent. Two point sources are simulated in the toy model, but only a single point source is included in the model.
While in the gradient descent method the best-fit can depend on the initialisation point (square symbols), the nested sampling method maps the full likelihood landscape and finds the multi-modal solution. The solid grey and red lines represent the trace plots of the \texttt{iminuit} gradient descent for two different initialisations. \github{https://github.com/gammapy/gammapy-2.0-paper/blob/main/scripts/nested_sampling_multimodal.py}}
\label{fig:cornerplot}
\end{figure}

To highlight the capacity of nested sampling to efficiently solve multi-modal problems with respect to the standard minimisation method (\codehref{https://scikit-hep.org/iminuit/}{iminuit}), we built the following 3D toy model. We simulated two point sources located at Galactic coordinates $(l,b)=(-0.10^\circ,-0.10^\circ)$ and
$(l,b)=(0.10^\circ,0.10^\circ)$, respectively. Both sources were assigned the same spectral model, a \codehref{https://docs.gammapy.org/2.0.2/api/gammapy.modeling.models.PowerLawSpectralModel.html}{PowerLawSpectralModel} with spectral index $\Gamma=2.0$ and an amplitude of $10^{-12}$\cmsTeV at 1\,TeV. 
We then utilised a model containing only one point source and performed both a standard fit and a nested sampling analysis. The expected solution is therefore multi-modal with one solution at each simulated position. The posterior distribution obtained with the \codehref{https://johannesbuchner.github.io/UltraNest/}{UltraNest} backend and the trace of the minimisation procedure are shown in Fig.~\ref{fig:cornerplot}. The trace plot illustrates the gradient-descent path taken by the \codehref{https://scikit-hep.org/iminuit/}{iminuit} algorithm \citep{James:1975dr, dembinski_2024_13923658}, showing that it converged to one of the two modes. Depending on the initialisation position, the fit converges to one of the two modes, whereas the nested sampling algorithm successfully identifies both modes. This simple example illustrates the strength of the nested sampling approach in the fact that it is robust to multi-modal solutions and is independent of the parameter initialisation, unlike gradient-descent or MCMC algorithms.

The toy model example shown here is intended to clearly illustrate the two different approaches. We do not explore all possible model parameters to test the robustness of the nested sampling. However, this method is known to be robust even in complex likelihood landscapes, such as the Gaussian shell, `Eggbox' \citep[see Fig.~6 and 7 of][]{Speagle2020} and diamond ring likelihood problems \citep[see Fig.~9 of][]{Buchner2023}. 
To conclude, there is no one-size-fits-all algorithm, and the choice of method should depend on the specific task at hand. Traditional gradient-descent fitting is rapid and works well for simple problems, but in higher-dimensional models it risks missing a mode or being stuck in a local minimum. Nested sampling provides a more robust global approach at the cost of requiring more likelihood evaluations.
The output products of both methods are also different and the choice depends on whether simple error estimate or computation of the full posterior distribution is needed for a given use case.

The \codehref{https://docs.gammapy.org/2.0.2/api/gammapy.modeling.Sampler.html}{Sampler} class interface is general enough that further backends can be implemented in the future. One such example is \codehref{https://nautilus-sampler.readthedocs.io}{Nautilus}, which uses likelihood emulators via neural networks for more efficient volume exploration in large dimension parameter space \citep{Lange2023}.

\section{Upper limits caveats and alternative approaches}
\label{section:upper_lims}

With the introduction of Bayesian inference in \texttt{Gammapy} v2.0, some caution is required when interpreting results, particularly for users accustomed to frequentist analyses. This is especially relevant for upper limit calculations, whose statistical interpretation and properties can differ substantially between the Bayesian and frequentist frameworks, as discussed in this section. 

The derivation of upper limits on physical parameters is a fundamental aspect of data analysis, but it comes with important caveats, particularly in Poisson-distributed count data such as \grays. The standard frequentist approach constructs a profile likelihood varying the parameter of interest and exploiting Wilks' theorem to relate the log-likelihood ratio to a chi-squared distribution, from which confidence intervals and one-sided upper limits at a desired confidence level (typically 95\% C.L.) are extracted \citep{Wilks1938,Cowan2012}. While this method is computationally efficient and well-motivated asymptotically, it carries several well-known limitations. 

In the context of \gray analysis, when the number of observed counts is sufficiently large, the upper limits derived by the likelihood profile method can provide a reasonable estimate. However, this method can become unreliable when the data contains few observed counts, due to short observation times (e.g. short timescale light curves), or at a very high energy (energies higher than tens of TeV).
Confidence-interval procedures, such as those of \citet{Feldman1998} and \citet{Rolke2005}, can provide upper limits with improved frequentist coverage, particularly in the low-count regime, where the asymptotic assumptions of the profile likelihood method may not accurately achieve the nominal confidence level. In such cases, dedicated simulations should be performed to verify the actual coverage of the method.

An alternative approach for upper limit derivation is Bayesian sampling, where the posterior distribution of samples is used to compute the parameter value for a given quantile (e.g. 95\%) of the samples. Importantly, this approach no longer relies on Wilks' theorem.
An advantage of this technique is that it naturally handles the nuisance parameters by marginalising over their posterior distribution and therefore propagates all systematic uncertainties into the confidence interval on the parameter of interest \citep[see e.g.][]{Gelman2013}. 

Frequentist confidence intervals are constructed to achieve a prescribed coverage probability under repeated sampling, whereas Bayesian credible intervals are defined as regions containing a specified fraction of the posterior probability. Consequently, Bayesian credible intervals do not carry a frequentist coverage interpretation and are not designed to satisfy a prescribed coverage criterion. The two approaches become asymptotically equivalent under suitable regularity conditions, such as those of the Bernstein-von Mises theorem \citep{vanderVaart1998}, including an approximately Gaussian likelihood, a true parameter value in the interior of the parameter space, and a locally non-informative prior. However, they can differ significantly when these conditions are not met, for example, when the parameter is constrained by a boundary (e.g. positivity constraint) or is in low-count regimes, where asymptotic approximations may break down. Such situations commonly arise in the computation of upper limits in \gray astronomy.

To illustrate the difference between the frequentist and Bayesian upper limits, we performed a toy model experiment in which we simulated a pure background dataset using CTAO South IRFs\footnote{\texttt{Prod5-South-20deg-AverageAz-14MSTs37SSTs.180000s-v0.1}} \citep{CTAO_irfs_2021}. For a given observation duration, we simulated the same dataset 1000 times to mimic a light-curve analysis of a non-detected transient follow-up. Observation durations of 0.03\,hours and 100\,hours were chosen to present datasets with both low counts ($\sim10$\,counts) and very high counts ($\sim3\times10^4$), respectively.
For each realisation, a spectral analysis between 50\,GeV and 50\,TeV was performed and upper limits of the amplitude of a power law with a fixed spectral index of 2.0 were derived with both methods.
The frequentist upper limits were derived using the profile likelihood method implemented in the \codehref{https://docs.gammapy.org/2.0.2/api/gammapy.estimators.FluxPointsEstimator.html}{FluxPointsEstimator} class. The Bayesian upper limits were derived using the sampling method implemented in the \codehref{https://docs.gammapy.org/2.0.2/api/gammapy.modeling.Sampler.html}{Sampler} class described in Sect.~\ref{section:sampling}. 
Two cases are demonstrated here, namely, one in which the flux is restricted to non-negative values (with a minimum flux of zero in the profile likelihood method and a uniform prior on the amplitude minimal and maximal values of [0, 10$^{-12}$]\cmsTeV in the sampling method) and an unbounded case in which negative amplitudes are allowed.

\begin{figure}
\centering
\includegraphics[width=\columnwidth]{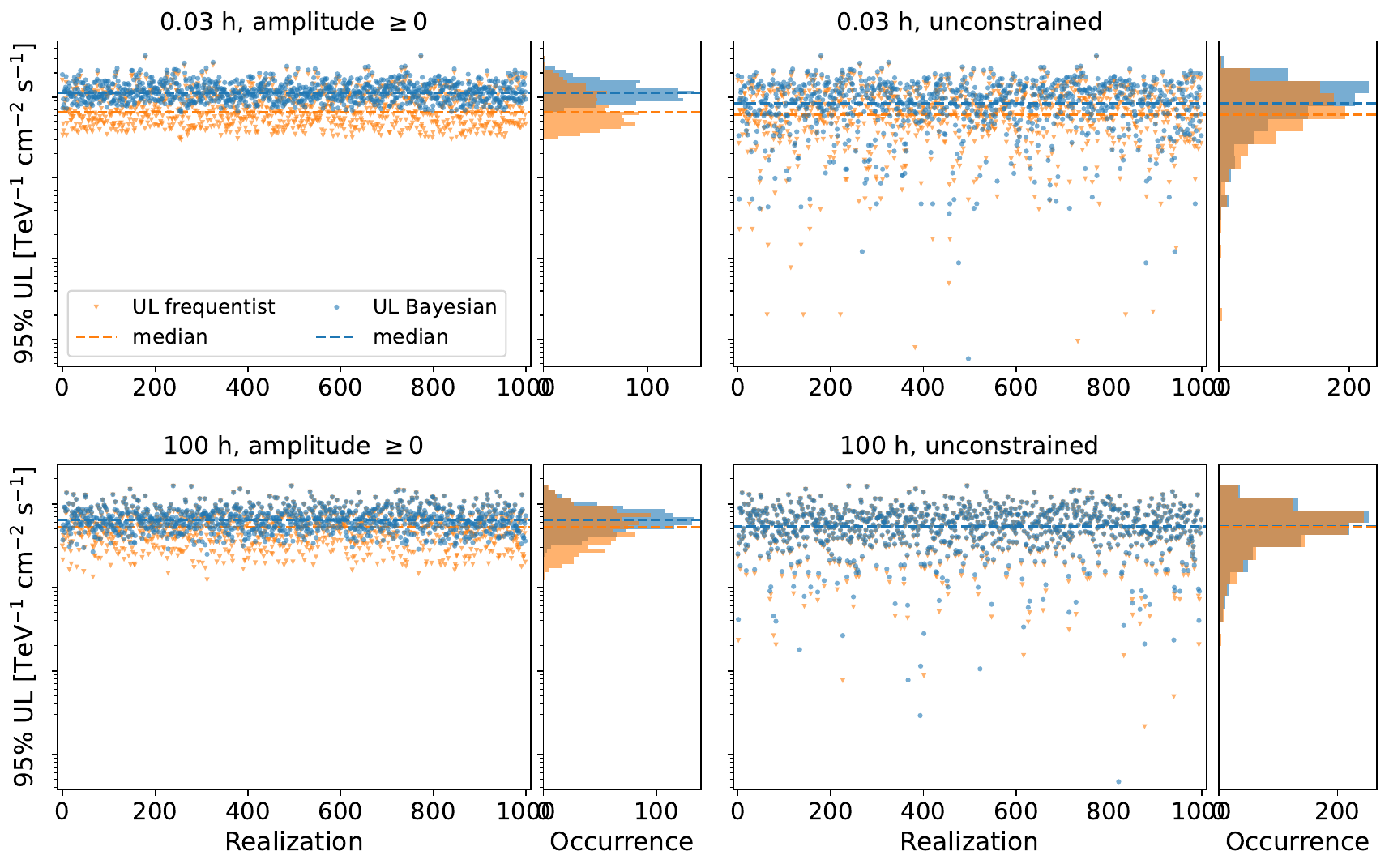}
\caption{Comparison of frequentist (profile likelihood, in orange) and Bayesian (sampling, in blue) 95\% upper limits for 1000 different random realisations of the same pure-background observation of CTAO-South. The top and bottom rows correspond to observation durations of 0.03\,h and 100\,h, respectively. In the left column, the amplitude is restricted to non-negative values, while the right column allows negative values (see main text for further detail). Each panel shows the upper limit for each realisation (left) and its corresponding distribution (right). Dashed lines indicate the medians. \github{https://github.com/gammapy/gammapy-2.0-paper/blob/main/scripts/UL_simulation_run.py} }
\label{fig:upper_limits}
\end{figure}

The resulting 95\% confidence upper limit (frequentist) and 95\% credible upper limit (Bayesian) are shown in Fig.~\ref{fig:upper_limits}. From the top to the bottom panel, one can see that the difference between the two quantities decreases as the observation time increases, approaching the asymptotic limit. However, in presence of a positivity constraint, the credible upper limit remains approximately 20\% higher than the confidence upper limit, even asymptotically (bottom-left panel). This is because, when the true parameter lies on the boundary, the posterior used to derive the credible upper limit (Bayesian) approaches a Gaussian distribution truncated at zero rather than a full Gaussian distribution. In contrast, the confidence upper limit (frequentist) remains unchanged under the positivity constraint because, although the latter modifies the asymptotic distribution, it yields the same one-sided 95\% critical value. The two quantities match only when the positivity constraint is relaxed and the observation time is sufficiently large, as seen in bottom-right panel of Fig.~\ref{fig:upper_limits}.

Overall, the new methods available in \texttt{Gammapy} v2.0 allow exploration of alternative ways, such as Bayesian sampling, to tackle the complex problem of upper limit derivation. Differences between Bayesian and frequentist upper limits should not be interpreted as indicating that one method is more accurate than the other. In general, the two approaches do not yield identical results because they address different statistical questions and quantify uncertainty in fundamentally different ways.
In future versions of \texttt{Gammapy}, we aim to provide further Bayesian analysis estimators as well as a method to quantify the robustness of both the frequentist and Bayesian upper limits.

\section{Sensitivity}
\label{section:sensitivity}

The sensitivity of a \gray instrument characterises the minimum flux that can be detected for a given source model with a specified statistical significance under pre-defined observational conditions. Traditionally, differential sensitivity is defined as the minimum source flux required to obtain a $5\sigma$ detection of a point-like source with a pre-defined spectrum (typically a power law with an index of two) in non-overlapping logarithmic energy bins, typically five per decade. In addition to the detection significance, further requirements are usually imposed on the minimum number of detected \gray events in each energy bin and on the signal-to-background ratio. This methodology is widely used in \gray astronomy and is implemented in the \codehref{https://docs.gammapy.org/2.0.2/api/gammapy.estimators.SensitivityEstimator.html}{SensitivityEstimator} class in \texttt{Gammapy} for one-dimensional spectral datasets with a Poisson background, following the significance estimation approach of \cite{LiMa1983}.

While this definition has proven useful for characterising instrument performance, it has several important limitations. In particular, it assumes an isolated point-like source with a one-dimensional spectrum and does not generalise to more complex analyses involving multiple event types or multiple overlapping sources. Such joint analyses are already employed in experiments such as HAWC and \fermi, which combine different event types, and are also foreseen for future instruments such as CTAO \citep{CTAConsortium:2023ynz}.

A more general statistical framework was proposed in Sect.~4 of \cite{Cowan2012} in which the sensitivity of a measurement is defined in terms of the expected median significance with which the null hypothesis can be rejected for a given assumed signal strength. In these more general settings, the median expected significance can be estimated without computationally expensive Monte Carlo simulations by using a representative artificial dataset known as the `Asimov dataset' \citep{Cowan2012}. 
This dataset is constructed with the pseudo-data counts set to the predicted counts for a given set of model parameter values instead of a Poisson-fluctuated realisation. Consequently, the significance obtained from the Asimov dataset directly gives the median expected significance. 

Sensitivity, in the context of \gray astronomy, is therefore the source flux for which the expected median significance, as given by the Asimov dataset, matches the detection threshold. A root finding algorithm can be used to determine this corresponding source flux.

\begin{figure}
\centering
\includegraphics[width=\columnwidth]{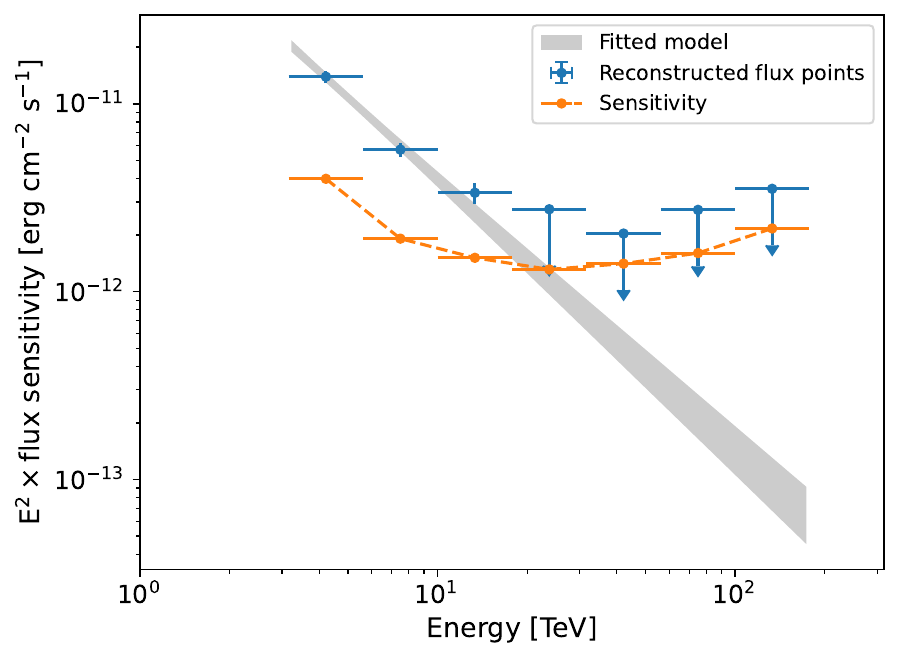}
\caption{Sensitivity (in orange) computed for an extended source with multiple event types using the Asimov dataset as described in the text. We considered an isolated extended source (extension of $\sigma=0.2^\circ$) and a power-law spectral index of 3.5, observed for 365 transits with the public HAWC IRFs with five different event types, available in \codehref{https://github.com/gammapy/gammapy-data}{GAMMAPY\_DATA}. The blue points show the reconstructed flux points and $5\sigma$ upper limits obtained from the simulated source. 
Note that sensitivity and upper limits represent different concepts. Sensitivity is an expected quantity that characterises the median detection capability of an experiment, whereas an upper limit is an observed quantity that provides the maximum signal flux compatible with a given confidence level for a specific dataset and is therefore subject to statistical fluctuations.
\github{https://github.com/gammapy/gammapy-2.0-paper/blob/main/scripts/sensitivity.py}}
\label{fig:sensitivity}
\end{figure}

In \texttt{Gammapy} v2.0, the \codehref{https://docs.gammapy.org/2.0.2/api/gammapy.estimators.FluxPointsEstimator.html}{FluxPointsEstimator} can be used to compute flux sensitivities following this approach. Since this works as a maximum-likelihood estimator with forward folding, this approach naturally supports joint analyses that enable different event types and modelling of spatial morphologies in crowded regions. As an example, we used the public HAWC IRFs supplied in \codehref{https://github.com/gammapy/gammapy-data}{GAMMAPY\_DATA}, which already include five different event types. We simulated an extended source with a Gaussian morphology (extension, $\sigma=0.2^\circ$) and spectral index of 3.5 with an integrated flux above 1\,TeV of $9.0\times 10^{-11}$\,erg\,cm$^{-2}$\,s$^{-1}$. In Fig.~\ref{fig:sensitivity}, we show the computed joint flux points and the sensitivity. Similarly, joint sensitivity maps can be computed in spatial coordinates using the \codehref{https://docs.gammapy.org/2.0.2/api/gammapy.estimators.TSMapEstimator.html}{TSMapEstimator}. It should be noted that these methods are only valid asymptotically and can provide incorrect estimates at very low statistics.

\section{Additional improvements}
\label{section:additional_improvements}

Beyond the major developments described previously, \texttt{Gammapy} v2.0 also includes a number of additional important improvements to the code base. We highlight a few of these additional enhancements in the following section.

\subsection{Energy-dependent morphology estimator}
As explained in detail in \cite{gammapy_v1}, the \codehref{https://docs.gammapy.org/2.0.2/api/gammapy.estimators.html}{gammapy.estimators} sub-module provides a range of algorithms for scientific analysis with a focus on hypothesis testing through the utilisation of the Wilks theorem \citep{Wilks1938}. 
A new methodology for measuring the energy-dependent morphology of a \gray source was introduced in the module \codehref{https://docs.gammapy.org/2.0.2/api/gammapy.estimators.EnergyDependentMorphologyEstimator.html}{EnergyDependentMorphologyEstimator}. This module is described in detail in a dedicated publication \citep{feijen_2025}, but we provide a brief overview here. 

This algorithm investigates the potential energy-dependent changes in the \gray source morphology in \texttt{Gammapy} by employing forward-folding methods with a 3D likelihood. It is able to quantify the significance of energy-dependent \gray morphology by performing a morphological fit of extension and position in various energy slices and comparing with a global morphology fit through Wilks theorem. This new method addresses key limitations of existing algorithms and offers a statistically robust framework for assessment.

\subsection{Timing utilities}

Since \texttt{Gammapy} v2.0, the timing-analysis framework has been significantly expanded to support variability studies. The \codehref{https://docs.gammapy.org/2.0.2/api/gammapy.modeling.models.LightCurveTemplateTemporalModel.html}{LightCurveTemplateTemporalModel} enables modelling of complex spectro-temporal phenomena with user-provided templates, including energy-dependent time delays, evolving flare profiles, and other complex temporal behaviour. Time-resolved spectral analysis can be performed directly on collections of \codehref{https://docs.gammapy.org/2.0.2/api/gammapy.datasets.Datasets.html}{Datasets}, facilitating background-subtracted variability studies within the standard analysis workflow.

Many standard variability estimation tools are now implemented within \texttt{Gammapy}, allowing users to directly quantify source variability from the computed light curves. These include measurements of fractional excess variance\footnote{\codehref{https://docs.gammapy.org/2.0.2/api/gammapy.estimators.utils.compute\_lightcurve\_fvar.html}{compute\_lightcurve\_fvar}} \citep{vaughan_2003}, flux doubling and halving times,\footnote{\codehref{https://docs.gammapy.org/2.0.2/api/gammapy.estimators.utils.compute\_lightcurve\_doublingtime.html}{compute\_lightcurve\_doublingtime}} and correlation-based analyses through the discrete correlation function\footnote{\codehref{https://docs.gammapy.org/2.0.2/api/gammapy.estimators.utils.compute_lightcurve_discrete_correlation.html}{compute\_lightcurve\_discrete\_correlation}} \citep{edelson_1988} and the \codehref{https://docs.gammapy.org/2.0.2/api/gammapy.stats.structure\_function.html}{structure\_function} \citep{emmanoulopoulos_2010}. The generation of synthetic light curves from a given power spectrum is now supported with the dedicated function\footnote{\codehref{https://docs.gammapy.org/2.0.2/api/gammapy.stats.TimmerKonig\_lightcurve\_simulator.html}{TimmerKonig\_lightcurve\_simulator}} \citep{timmer_1995}.
Together, these tools enable investigations of variability amplitudes, characteristic timescales, and temporal correlations across multiple energy bands. 

Pulsar analysis capabilities have matured through enhancements to the \codehref{https://docs.gammapy.org/2.0.2/api/gammapy.makers.PhaseBackgroundMaker.html}{PhaseBackgroundMaker}, which now correctly propagates phase intervals and supports energy-dependent radial cuts, improving the signal-to-noise ratio. The \codehref{https://docs.gammapy.org/2.0.2/api/gammapy.modeling.models.TemplatePhaseCurveTemporalModel.html}{TemplatePhaseCurveTemporalModel} has been adapted to support both pulsar and binary star analysis.
Additional simulation and fitting interfaces in the temporal domain may be found in the external package \codehref{https://github.com/cgalelli/gammapy_SyLC}{Gammapy-SyLC} \citep{galelli_2025}.

\subsection{Hypotheses testing for nested models}

An important new feature that allows the user to easily calculate the test statistic between two nested hypotheses has been introduced with the \codehref{https://docs.gammapy.org/2.0.2/api/gammapy.modeling.select_nested_models.html}{select\_nested\_models} function. In this context, a nested model is a model for which one hypothesis can be obtained from the other by fixing one or more additional parameters to specific values such that the simpler model constitutes a limiting case of the more general one. For example, a point source is a limiting case of an extended source, with its extension parameter equal to zero.
The functionality assesses whether the inclusion of one or more additional parameters, corresponding to the alternative hypothesis, yields a statistically significant improvement over the null hypothesis, in which case the parameters are fixed at specified values. The model is updated to reflect the alternative hypothesis if the observed improvement exceeds a pre-defined significance threshold. 
This can be used in a number of science use cases, for example, to determine whether the addition of a source to the model is significant or to test for specific features in a model (e.g. test the existence of a spectral cutoff) and automatically select the preferred models (e.g. while making catalogues). 

To explain the use of \codehref{https://docs.gammapy.org/2.0.2/api/gammapy.modeling.select_nested_models.html}{select\_nested\_models}, we show how to test if a given source is statistically significant with the code snippet in Fig.~\ref{fig:snm_code_snippet}. This function performs an internal comparison between nested models corresponding to the null and alternative hypotheses. In this context, the null hypothesis represents the absence of the source, while the alternative hypothesis represents the presence of a significant source. Therefore, under the null hypothesis, the amplitude of the source is fixed to zero, corresponding to a flux of zero. Before calling the function, the spatial parameters of the source are held fixed to prevent the fit from converging to unrelated positions. This also prevents a weaker nearby source from shifting its position to replace the source under investigation during the null-hypothesis fit. 

\begin{figure}[htbp]
\begin{lstlisting}
from gammapy.modeling.selection import (
    select_nested_models
)
from gammapy.datasets import (
    Datasets,
    SpectrumDatasetOnOff,
)
from gammapy.modeling.models import SkyModel

path = "$GAMMAPY_DATA/fermi-3fhl-crab/"
filename = f"{path}Fermi-LAT-3FHL_datasets.yaml"
filename_models = f"{path}Fermi-LAT-3FHL_models.yaml"
fermi_datasets = Datasets.read(
    filename=filename,
    filename_models=filename_models
)

model = fermi_datasets.models["Crab Nebula"]

# number of free parameters in the model
n_free_params = len(model.parameters.free_parameters)

# Manually prevent source from moving position
fermi_datasets.models.freeze(model_type="spatial")

# test if source is significant
results = select_nested_models(
    fermi_datasets,
    parameters=[model.spectral_model.amplitude],
    null_values=[0],
    n_free_parameters=n_free_params,
    n_sigma=4,
)
\end{lstlisting}
\caption{Code snippet of the \texttt{select\_nested\_models} that tests if a source is detectable above 4$\sigma$. The null hypothesis (no source i.e. $\mathrm{amplitude}=0$) is selected otherwise.}
\label{fig:snm_code_snippet}
\end{figure}

\subsection{Development and operation practices}
This section outlines the new additions to the DevOps of our code base, which is a combination of development (Dev) and operations (Ops) practices. The appearance of the \texttt{Gammapy} documentation was enhanced by utilising the PyData Sphinx theme directly, allowing for improved accessibility of the pages, including support for both the light and dark themes. A proper deprecation system was introduced as inspired by the \texttt{Astropy} package \citep{astropy_2022}, allowing us to appropriately handle application programming interface (API) breaking changes. Through the use of decorators, a warning is raised when a deprecated feature is called, allowing users to be warned of future changes affecting their code. These changes are also clearly outlined in the release notes. This is an important addition, as we are now in the major stages of the \texttt{Gammapy} software release. We now distribute wheels on PyPI utilising the OpenAstronomy release workflow, which provides easy installation on various architectures. 

An effort was made to replace the \texttt{Cython} code with just in time (JIT) compiled code, resulting in a speed-up in performance using the \codehref{https://numba.pydata.org/}{Numba} JIT compiler \citep{numba}. Currently both approaches are implemented for testing purposes. In the longer term, the aim is to retain only the JIT-compiled code, thus providing a pure Python replacement of the \texttt{Cython} code.

The addition of the pre-commit continuous integration (CI) and the \codehref{https://docs.astral.sh/ruff/}{ruff} linter and formatter became increasingly important for adhering to formatting guidelines and coding standards and aligning with other well-established Python packages. This feature automatically detects syntax errors and formatting issues, and it enforces specific coding standards. In line with this, we also exhaustively tested all code in the reStructuredText (RST) documentation files and docstrings to ensure no code failures and correct syntax.

\section{Software growth and community engagement}
\label{section:community_software}

\begin{figure}[hbt]
\centering
\includegraphics[width=\columnwidth]{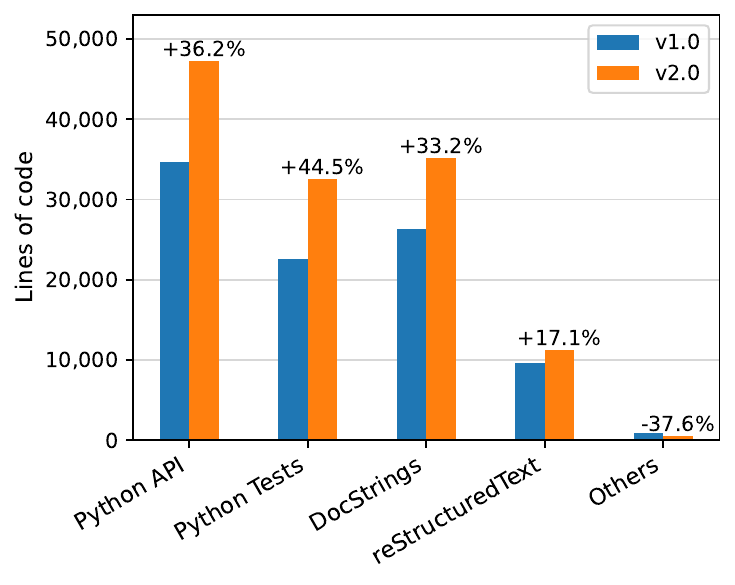}
\caption{Overview of how the number of lines of code are distributed in v1.0 vs v2.0, illustrating the growth of the code base. \github{https://github.com/gammapy/gammapy-2.0-paper/blob/main/scripts/lines_of_code_calculation.py}}
\label{fig:loc_type}
\end{figure}

\texttt{Gammapy} is widely used by the astronomy community, with 352 citations\footnote{This is obtained from an ADS script taking unique citations from Zenodo and referred papers.} at the date of writing. Since the first long-term stable release, version 1.0, the code base has undergone substantial development, including a significant increase in the total number of lines of code. A comparison between v1.0 and v2.0 for the number of lines of code in the software codebase is shown in Fig.~\ref{fig:loc_type}.

In addition to the code development, the project team has placed a strong emphasis on improving the documentation quality. This was made possible following extensive and important feedback from both users and developers, in particular from the \texttt{Gammapy} coding sprints\footnote{\url{https://github.com/gammapy/gammapy-meetings/tree/master/coding-sprints}} and the schools we are organising.\footnote{e.g. \url{https://indico.obspm.fr/event/2672/}} The documentation appearance was upgraded, and the content was revised and expanded, leading to clearer guides and explanations and improved accessibility for both new and experienced users. 
In particular, the number of tutorials has increased by 42\%, from 36 to 51, providing more comprehensive examples to support the understanding and effective use of the software.
This is intended to help reduce the steepness of the learning curve for beginners.

The engagement of the community has also grown and evolved between the two long-term stable releases. As a comparison, the number of contributors based on the corresponding authors for the relevant Zenodo releases shows a 15\% increase, from 44 to 51 contributors. In fact, since v1.0 we have had 35 new contributors to the package. In addition, the lead developers team has evolved and expanded into a larger and more balanced team. These positive developments highlight the importance of maintaining an community-driven open-source software project.

\begin{figure}[hbt]
\centering
\includegraphics[width=\linewidth]{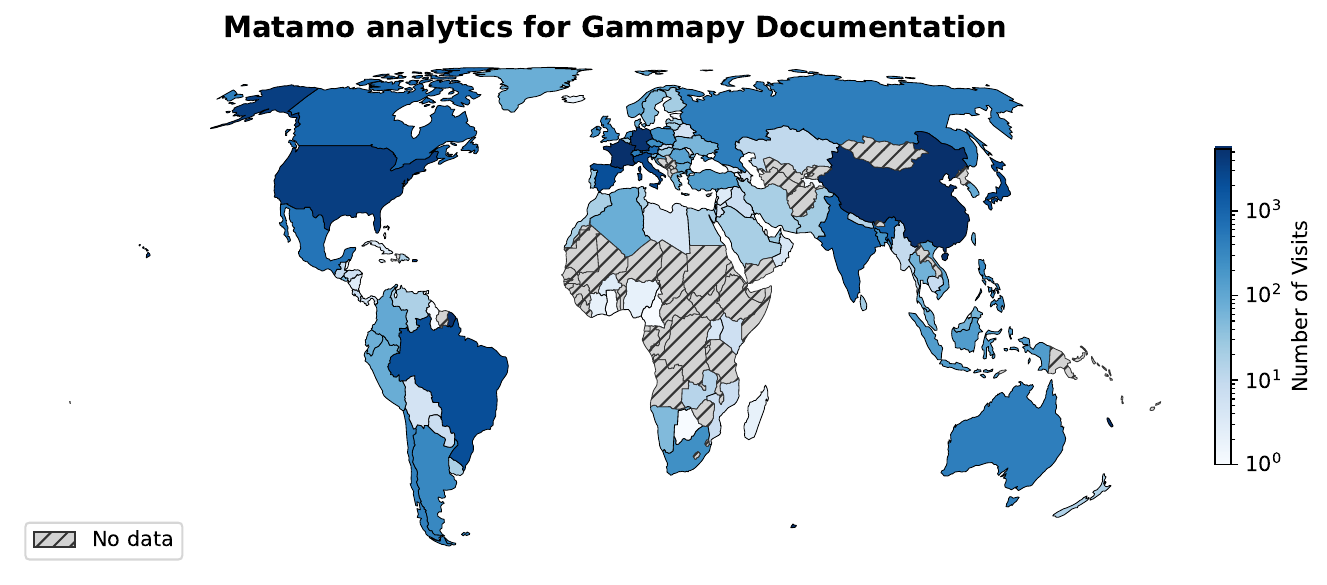}
\caption{Number of visits per country who accessed the \texttt{Gammapy} documentation from September 2024 until February 2026\protect\footnotemark. \github{https://github.com/gammapy/gammapy-2.0-paper/blob/main/scripts/visitors_map.py}}
\label{fig:visits_map}
\end{figure}

The \texttt{Gammapy} software is utilised by a considerable number of users worldwide, as showcased by the number of visits to the \texttt{Gammapy} documentation between September 2024 and February 2026 in Fig.~\ref{fig:visits_map}. In 2024, Matomo Analytics\footnote{\url{https://matomo.org/}} was introduced, thus replacing Google Analytics with an open-source web analytics platform, to monitor the use of the documentation webpage. These statistics enable us to gain an understanding of the geographical distribution and diversity of the \texttt{Gammapy} community, including its users and contributors. 

\footnotetext{Note: The country-level boundaries shown in this map are provided for statistical visualisation purposes only. They do not imply any position by the authors or their respective institutions regarding the legal status or delimitation of any frontiers.}

The widespread use of \texttt{Gammapy} is further highlighted by the increasing number of experiments that have validated their data analyses with \texttt{Gammapy}, which was not the case at the time of \texttt{Gammapy} v1.0. These include, but are not limited to, MAGIC \citep{2024_magic_validation}, VERITAS \citep{veritas_2023}, ASTRI \citep{astri_2022}, and LST-1 \citep{LST_2023}. 
Furthermore, \texttt{Gammapy} is featured among the high-impact open-source software projects identified in a recent survey of scientific software development in astronomy, reflecting its substantial growth and its impact \citep{buchner_impact_2025}.

\section{Conclusions}
\label{conclusions}

This paper has presented \texttt{Gammapy} v2.0, the second long-term stable release of the \texttt{Gammapy} open-source Python package for high-level \gray data analysis. This release builds on the architecture and concepts described in the first \texttt{Gammapy} paper \citep{gammapy_v1} while consolidating and extending functionality developed over the past several years. The major new features include support for Bayesian analyses via nested sampling, the computation of joint TS maps and combined significance maps, the computation of joint sensitivity, flux points and maps, an estimator to study energy-dependent morphologies, and several timing utilities. The importance of a community-driven open-source software is highlighted by the widespread contributions to the \texttt{Gammapy} project managed by a team of core developers.

By relying on the GADF standard, \texttt{Gammapy} can process high-level data products from a variety of \gray instruments, including imaging atmospheric Cherenkov telescopes, water Cherenkov detectors, and space telescopes. This design choice allows \texttt{Gammapy} to remain largely agnostic to instrument specifics and facilitates its use in the context of current and future open observatories. Standardisation has only recently started within the current generation of instruments but is progressing rapidly with support for \fermi, \hess \citep{2019_mohrmann_hess}, HAWC \citep{2022_hawc_validation}, MAGIC \citep{2024_magic_validation}, and VERITAS \citep{wong_2025_15092381}. In the future, \texttt{Gammapy} will be the default for the next generation of \gray instruments such as CTAO and SWGO, and it can also be extended to support neutrino data from experiments such as KM3Net and IceCUBE. 
In this landscape, there is an increased need for multi-wavelength and multi-messenger analyses, which motivates the development of the shared open data format for very-high-energy \gray and neutrino astronomy known as the Very-high-energy Open Data Format, or VODF \citep[][]{vodf}. We aim for better support of systematic uncertainties in future releases; however, with the existing tools, it is already possible to gain an understanding of these uncertainties, as shown by \cite{nigro_2019}.

The introduction of the \codehref{https://github.com/gammapy/gammapy-mwl/}{gammapy-mwl} sub-package extends the scope of this framework beyond \gray data alone, providing initial support for multi-wavelength analyses, in particular for X-ray observations. This represents a step towards integrated multi-instrument workflows that can be handled within a common analysis environment, allowing multi-wavelength forward-folding modelling from eV to TeV energies as presented in \citet{NievasRosillo2025}. 

Development towards the next major release will focus on the consolidation and extension of support for analyses relevant to the Cherenkov Telescope Array Observatory, with the integration of \texttt{Gammapy} into the CTAO SAT package. Continued efforts are also planned to improve software sustainability through DevOps-related developments, including refinements of continuous integration, testing strategies, and release workflows, in line with CTAO software requirements. 

Further work will address metadata handling and the evolution of data formats, with the aim of improving interoperability, validation, and long-term usability of high-level data products. In addition, planned developments include the extension of Bayesian analysis functionalities, a more versatile high-level interface, support for unbinned analyses, support for GPU acceleration, data access through the Virtual Observatory \citep{heig2024}, improvements to features associated with dark-matter analyses, and further enhancements of the statistical tools available in the framework. These topics emerged from feature requests and discussions in the \texttt{Gammapy} issue tracker, slack channels, and user calls, which are largely driven by community input.

\begin{acknowledgements}
The \texttt{Gammapy} team acknowledges all past and current contributors to the library, as well as all contributors of the main \texttt{Gammapy} dependency libraries: 
\codehref{https://numpy.org/}{NumPy} \citep{harris2020array}, \codehref{https://scipy.org/}{SciPy} \citep{2020SciPy-NMeth}, \codehref{http://www.astropy.org}{Astropy} \citep{astropy_2022}, \codehref{https://astropy-regions.readthedocs.io/}{Astropy Regions}, \codehref{https://scikit-hep.org/iminuit/}{iminuit} \citep{James:1975dr, dembinski_2024_13923658}, and \codehref{https://matplotlib.org/}{Matplotlib} \citep{Hunter:2007}.
We would like to thank all the Gammapy users that have provided feedback and submitted bug reports. 
The authors acknowledge the \href{https://www.acme-astro.eu/}{ACME} project, which has received funding from the European Commission’s Horizon Europe Research and Innovation programme under grant agreement No. 101131928. They also acknowledge the \href{https://oscars-project.eu/}{OSCARS} project, which has received funding from the European Commission’s Horizon Europe Research and Innovation programme under grant agreement No. 101129751. 
For contributing to the writing of the manuscript, we thank F.~Acero, T.~Bylund, K.~Feijen, C.~Galelli, B.~Khéllifi, Q.~Remy, A.~Sinha, and R.~Terrier
\end{acknowledgements}

\bibliographystyle{aa}
\bibliography{bibliography}

\end{document}